\documentclass[preprint,review,3p,12pt]{elsarticle}

\usepackage{amssymb}

\journal{Journal of Crystal Growth}

\begin{document}

\begin{frontmatter}



\title{Single crystal growth of the pyrochlores $R_2$Ti$_2$O$_7$ ($R$ = rare earth)
by the optical floating-zone method}


\author{Q. J. Li}
\author{L. M. Xu}
\author{C. Fan}
\author{F. B. Zhang}
\author{Y. Y. Lv}
\author{B. Ni}
\author{Z. Y. Zhao}
\author{X. F. Sun\corref{cor1}}


\address{Hefei National Laboratory for Physical Sciences at Microscale,
University of Science and Technology of China, Hefei, Anhui
230026, China}

\cortext[cor1]{Corresponding author. Tel.: 86-551-63600499, Fax:
86-551-63600499.\\ Email address: xfsun@ustc.edu.cn}


\date{\today}

\begin{abstract}

We report a systematic study on the crystal growth of the
rare-earth titanates $R_2$Ti$_2$O$_7$ ($R$ = Gd, Tb, Dy, Ho, Y,
Er, Yb and Lu) and Y-doped Tb$_{2-x}$Y$_x$Ti$_2$O$_7$ ($x$ = 0.2 and
1) using an optical floating-zone method. High-quality single
crystals were successfully obtained and the growth conditions were
carefully optimized. The oxygen pressure was found to be the most
important parameter and the appropriate ones are 0.1--0.4 MPa,
depending on the radius of rare-earth ions. The growth rate is
another parameter and was found to be 2.5--4 mm/h for different
rare-earth ions. X-ray diffraction data demonstrated the good
crystallinity of these crystals. The basic physical properties of
these crystals were characterized by the magnetic susceptibility
and specific heat measurements.

\end{abstract}


\end{frontmatter}



\newpage

\section{Introduction}

The rare-earth titanate with the pyrochlore structure,
$R_2$Ti$_2$O$_7$ ($R$ = rare earth), crystallizes into a face
centered cubic structure with eight formula units in a unit cell,
and the space group is $\emph{Fd$\overline{3}m$}$. The rare earth
ions form a network of corner-sharing tetrahedra, of which the
vertices are occupied by $R^{3+}$, with triangular and kagom\'{e}
planes alternately stacked along the [111] direction. Recently,
these materials have attracted considerable interests because of
their three-dimensional feature of geometrical magnetic
frustration and the resulting exotic ground states at low
temperatures, including spin ice, spin liquid and
order-by-disorder \cite{Structure, Spin_liquid, Spin_ice,
Er2Ti2O7_SH}. Since these low-temperature magnetic properties are
sometimes very sensitive to the sample quality, as was observed in
Yb$_2$Ti$_2$O$_7$ and Tb$_2$Ti$_2$O$_7$ \cite{Yb2Ti2O7_1,
Tb2Ti2O7_1}, the growth of high-quality $R_2$Ti$_2$O$_7$ single
crystals is very important for investigating the intrinsic physics
of these materials. Both the flux method and the optical
floating-zone technique have been tried for growing
$R_2$Ti$_2$O$_7$ single crystals and the latter was found to be
able to produce better and bigger crystals \cite{flux, FZ,
Tb2Ti2O7, Y2Ti2O7, Dy2Ti2O7}. However, the detailed growing
conditions of $R_2$Ti$_2$O$_7$ using the floating-zone method have
not been systemically studied or reported and the crystal quality
demonstrated in the literature seems to have significant room to
improve.

In this work, we studied in details the conditions for the crystal
growth of the $R_2$Ti$_2$O$_7$ ($R$ = Gd, Tb, Dy, Ho, Y, Er, Yb
and Lu) using the floating-zone method. It was found that the
growth rate, atmosphere and oxygen pressure should be carefully
adjusted to get high-quality single crystals and they differ from
each other for different rare-earth ions. Moreover, to study the
magnetic dilution effect of the spin liquid Tb$_2$Ti$_2$O$_7$, we
have grown Tb$_{2-x}$Y$_x$Ti$_2$O$_7$ ($x$ = 0.2 and 1) single
crystals by replacing magnetic Tb$^{3+}$ ions with nonmagnetic
Y$^{3+}$ ions. The obtained crystals were well characterized by
X-ray diffraction, magnetic susceptibility and specific heat
measurements.

\section{Crystal Growth}

Single crystals of $R_2$Ti$_2$O$_7$ ($R$ = Gd, Tb, Dy, Ho, Y, Er,
Yb and Lu) and Tb$_{2-x}$Y$_x$Ti$_2$O$_7$ ($x$ = 0.2 and 1) were
grown using an optical floating-zone furnace with four 1000 W
halogen lamps (Crystal System Incorporation, Japan).
Dy$_2$Ti$_2$O$_7$ was chosen as a trial to find the optimum growth
condition for pyrochlore titanates. There are some earlier reports
that Dy$_2$Ti$_2$O$_7$ single crystals could be grown in different
atmospheres (O$_2$, Ar+O$_2$ or Ar) with different growth rates
\cite{FZ, Dy2Ti2O7}. We found that the quality of
Dy$_2$Ti$_2$O$_7$ single crystals is quite sensitive to such
growing conditions as oxygen pressure and growth rate. At the
beginning, we tried the growth at a rate of 2.5 mm/h and in
flowing oxygen with 0.3 MPa pressure, which can ensure a stable
growing. The diameter of crystal is very uniform compared to that
grown in air (ambient pressure) in an earlier report \cite{FZ}.
The single crystal bar is transparent with the color of amber and
develops small facets. However, it has some small cracks at the
end part, as shown in Fig. 1(a), indicating that the growth
condition is still not appropriate. Then, we tried to increase the
oxygen pressure to 0.4 MPa and grew it with a rate of 2 mm/h, in
which case the molten zone could not be stabilized and it dropped
when the crystal growth has been continued for only 3--4 hours. A
possible improving way is to change the growth rate. After several
attempts, we finally got the optimum condition of 0.4 MPa oxygen
pressure and 4 mm/h rate for Dy$_2$Ti$_2$O$_7$. As shown in Fig.
1(b), the as-grown single crystal displays a nice morphology, long
consecutive facets and a homogenous amber color. The powder X-ray
diffraction measured on a randomly selected part of
Dy$_2$Ti$_2$O$_7$ single crystal confirmed the pure and single
phase. In addition, a narrow width of the rocking curve of the
(440) Bragg peak (FWHM $\approx 0.11^\circ$), as shown in Fig. 2,
demonstrates that the crystal has good crystallinity. Therefore,
the appropriate oxygen pressure and growth rate play important
roles in the growth of Dy$_2$Ti$_2$O$_7$ single crystals.

Similar investigations have been done for growing other
$R_2$Ti$_2$O$_7$ crystals and the obtained single crystals are
presented in Fig. 3. It was found that Ho$_2$Ti$_2$O$_7$ and
Gd$_2$Ti$_2$O$_7$ could also be grown under 0.4 MPa oxygen
pressure while the growth rate is another adjustable parameter,
which is important for stabilizing the molten zone \cite{FZ}. A
homogenous high-quality Ho$_2$Ti$_2$O$_7$ crystal could be
obtained with a rate of 4 mm/h. A lower speed of 2.5 mm/h was
found to be better for growing Gd$_2$Ti$_2$O$_7$. On the other
hand, the growth rate was found to be less crucial for
Tb$_2$Ti$_2$O$_7$ under 0.4 MPa oxygen pressure; high-quality
crystals were grown with the rate of either 2.5 or 4 mm/h. The
situations are a bit different for Er$_2$Ti$_2$O$_7$ and
Yb$_2$Ti$_2$O$_7$, it is found that the crystals grown under high
oxygen pressure always have a large number of small cracks even
though different growth rates were tried. Very-high-quality single
crystals of Er$_2$Ti$_2$O$_7$ and Yb$_2$Ti$_2$O$_7$ can be grown
only under 0.1 MPa oxygen pressure with the rate of 4 and 3 mm/h,
respectively. Whereas Y$_2$Ti$_2$O$_7$ requires even smaller
oxygen pressure and our best conditions for this crystal are in a
normal pressure oxygen-argon mixture with the ratio of 4:1 and at
a rate of 4 mm/h. The obtained Y$_2$Ti$_2$O$_7$ still has some
small cracks at the end part, which might be improved by some
slight modification of the growth condition. In addition, the
single crystal of Lu$_{2}$Ti$_2$O$_7$ was obtained under 0.1 MPa
oxygen pressure with rate of 4.5 mm/h, and has a morphology
similar to that of Y$_2$Ti$_2$O$_7$. Single crystals of
Tb$_{1.8}$Y$_{0.2}$Ti$_2$O$_7$ and TbYTi$_2$O$_7$ were
successfully obtained under 0.25 MPa oxygen pressure with rates of
2.5 and 3 mm/h, respectively, and have similar appearance to that
of Tb$_2$Ti$_2$O$_7$. The optimum growth conditions of the
$R_2$Ti$_2$O$_7$ single crystals are included in Table 1. The
lattice constants are obtained from the powder X-ray diffraction
data and are found to be consistent with the results from the
literature \cite{parameter}. It should be noted that the lattice
constants of $R_2$Ti$_2$O$_7$ have some weak dependencies on the
oxygen content \cite{Tb2Ti2O7, Dy2Ti2O7}. In our case, the lattice
constants of the Tb$_2$Ti$_2$O$_7$ crystals grown in 0.4 MPa
oxygen pressure and annealed in 0.1 MPa oxygen pressure are
10.155(2) and 10.152(1){\AA}, respectively, which also indicates
the weak dependence of lattice constant on oxygen content.

It is worth noting that the quality of $R_2$Ti$_2$O$_7$ crystals
can be easily judged from their appearances. Our subsequent
structural and physical-property analyses demonstrated that
crystals with good appearances, like shinning surface, homogenous
color and no cracks, always have high quality. The less successful
growths of single crystals with poor appearances were found to
have impurity phases. For example, the X-ray diffraction indicated
that the single crystal Dy$_2$Ti$_2$O$_7$ obtained with a rate of
2 mm/h has a impurity phase of Dy$_2$TiO$_5$. On the other hand,
the color change of the $R_2$Ti$_2$O$_7$ single crystals could be
indicative of slight variation in the oxygen content
\cite{Dy2Ti2O7}. As shown in Figs. 1 (a) and 1 (b), the color of
Dy$_2$Ti$_2$O$_7$ single crystal grown in 0.4 MPa oxygen pressure
is deeper than that grown in 0.3 MPa oxygen pressure. Apparently,
the crystal with lighter color has smaller oxygen content. A
post-annealing confirmed this, that is, the color of
Dy$_2$Ti$_2$O$_7$ single crystal becomes much lighter after
annealing the crystal in normal-pressure O$_2$, as shown in Fig. 1
(c). This phenomenon demonstrates that deeper-colored
Dy$_2$Ti$_2$O$_7$ crystal has larger oxygen content and is more
likely free from the oxygen vacancies. Similar phenomenon was also
found in other $R_2$Ti$_2$O$_7$ crystals.

One remarkable problem which has been encountered for the growth
of $R_2$Ti$_2$O$_7$ single crystals is how to avoid small cracks.
It has been discussed in previous reports that single crystal was
easily to have thermal cracks if a single-crystal seed was used
\cite{Dy2Ti2O7}. However, we found that seed bar does not have any
influence on the quality of single crystal; in contrast, suitable
oxygen pressure and growth rate are more important to reduce the
thermal cracks and get good single crystals.

The most important finding in the present work is that the optimum
growth conditions are obviously different for different rare-earth
titanates. As displayed in Table 1, those compounds with
relatively larger radius of $R^{3+}$ ions need to be grown in
higher oxygen pressure, like Gd$_2$Ti$_2$O$_7$, Tb$_2$Ti$_2$O$_7$,
Dy$_2$Ti$_2$O$_7$, Ho$_2$Ti$_2$O$_7$, while those with small
radius of $R^{3+}$ ions should be grown under relatively low
pressure, such as Er$_2$Ti$_2$O$_7$, Yb$_2$Ti$_2$O$_7$,
Y$_2$Ti$_2$O$_7$ and Lu$_2$Ti$_2$O$_7$. The growth condition for
Tb$_{1-x}$Y$_{x}$Ti$_2$O$_7$ single crystals also confirms this
law. As we known, pyrochlore oxides are represented by the
chemical formula $A_2B_2$O$_6$O$'$, where $A$ is a trivalent rare
earth consisting of the lanthanides, Y, or Sc, and $B$ could be a
transition metal, which is sitting at $16c$ and $16d$ sites of the
$\emph{Fd$\overline{3}m$}$ space group, respectively. Both $A$ and
$B$ form three-dimensional corner-sharing tetrahedra and are
coordinated by oxygen ions, and the content of the O$'$ atom plays
an important role in the coordination geometry of the $A$ and $B$
sites \cite{Structure}. In addition, the structure-field or
stability-field map for $A_2B_2$O$_7$ materials show that not only
the stability of pyrochlore phase but also the defect
concentration are influenced by the radius ratio of
$A^{3+}$/$B^{4+}$. For instance, the pyrochlore phase must be
synthesized by using high pressure if $B$ ion has very small
radius, like Mn$^{4+}$ ion \cite{Structure}. It is therefore
understandable that the oxygen pressure plays a crucial role in
the growth of $R_2$Ti$_2$O$_7$ crystals and avoiding thermal
cracks. One may note that in the present work the oxygen pressure
for the optimized growth is larger than those in earlier reports
\cite{FZ, Dy2Ti2O7}. The main reason why a high oxygen pressure is
helpful for growing $R_2$Ti$_2$O$_7$ is that it can reduce the
evaporation from molten zone and suppress the micro bubbles at the
melting interface. Furthermore, a high oxygen pressure could avoid
the oxygen deficiency and stabilize the phase formation of
$R_2$Ti$_2$O$_7$.

\section{Magnetic susceptibility and specific heat}

Besides the structural characterizations of the obtained
$R_2$Ti$_2$O$_7$ and Tb$_{2-x}$Y$_x$Ti$_2$O$_7$ single crystals
using powder X-ray diffraction, single-crystal rocking curve and
Laue photograph, the basic physical properties of these crystals
were also characterized. DC magnetization and specific heat
measurements were done using a SQUID-VSM (Quantum Design) and a
Physical Property Measurement System (PPMS, Quantum Design),
respectively. In general, both the magnetic susceptibility and
specific heat results of our single crystals are consistent with
most of data in literatures. Here we show some representative
data.

Fig. 4 shows the magnetization curves of Gd$_2$Ti$_2$O$_7$ and
Er$_2$Ti$_2$O$_7$ at 2 K along three characteristic axes [100],
[111], and [110]. Since the demagnetization effect is not
negligible in this system, the magnetization measurements along
three axes are done on samples with similar long-bar shape, of
which the size is about 2$\times$0.75$\times$0.75 mm$^3$. The
magnetic field is always applied along the longest dimension,
therefore the demagnetization factor is about 0.15 for these
samples. Gd$_2$Ti$_2$O$_7$ is regarded as a Heisenberg
antiferromagnet since Gd$^{3+}$ has no orbital moment. But
actually, it exhibits weak anisotropy at low temperatures as shown
in Fig. 4(a), which means that other exchange interactions than a
simple Heisenberg one should be taken into account
\cite{Gd2Ti2O7_SH, Gd2Ti2O7_MH}. Although the specific heat,
neutron scattering and some other experiments indicated that
Er$_2$Ti$_2$O$_7$ has a strong local {\it XY}-type anisotropy
\cite{Er2Ti2O7_SH, Er2Ti2O7}, there has been no report on the
anisotropic magnetization of Er$_2$Ti$_2$O$_7$ single crystals.
Our magnetization results in Fig. 4(b) show a weak anisotropy of
Er$_2$Ti$_2$O$_7$ at 2 K. In addition, the magnetic
susceptibilities of Tb$_2$Ti$_2$O$_7$ crystals annealed under 0.1
MPa oxygen pressure were measured and compared with that of
as-grown crystals. It was found that the magnetism down to 2 K is
almost independent of the oxygen content and just shows subtle
difference, as shown in Fig. 5 \cite{Tb2Ti2O7}.

Low-temperature specific heat data of $R_2$Ti$_2$O$_7$ ($R$ = Dy,
Tb, Gd, Er, Yb, Y and Lu) single crystals are shown in Fig. 6(a).
The nonmagnetic Y$_2$Ti$_2$O$_7$ and Lu$_2$Ti$_2$O$_7$ show a
simple behavior of lattice heat capacity. The specific heat of
Dy$_2$Ti$_2$O$_7$ exhibits a peak around 1 K but does not show any
sign of long-range order down to 0.4 K \cite{Dy2Ti2O7_SH}. There
is only one broad peak around 2.5 K in the specific-heat data of
Yb$_2$Ti$_2$O$_7$ with temperature down to 0.4 K
\cite{Yb2Ti2O7_SH1, Yb2Ti2O7_SH2}. For Gd$_2$Ti$_2$O$_7$, there
are two sharp peaks at 0.7 and 1 K, which are attributed to the
development of long-range magnetic order below about 1 K
\cite{Gd2Ti2O7_SH, Gd2Ti2O7_SH1}. The specific heat of
Er$_2$Ti$_2$O$_7$ also displays a sharp peak at about 1.2 K,
corresponding to a second-order phase transition
\cite{Er2Ti2O7_SH, Er2Ti2O7_SH1}. Two broad peaks at about 0.7 and
6 K are observed in the specific-heat data of Tb$_2$Ti$_2$O$_7$.
It is fundamentally consistent with previous reports
\cite{Tb2Ti2O7_1, Tb2Ti2O7_SH1, Tb2Ti2O7_SH2, Tb2Ti2O7_SH3},
although there are some differences of the data among different
samples. The peak at 6 K was attributed to a remnant of an
excitation between the ground state doublet and an excited
doublet, separated by $\sim$ 18 K, which was related to the
short-range magnetic correlation \cite{Tb2Ti2O7_SH1,
Tb2Ti2O7_SH2}. The peak at 0.7 K was mainly attributed to the
splitting of the ground state doublet, and the same short-range
magnetic correlation effect also can not be neglected
\cite{Tb2Ti2O7_SH1, Tb2Ti2O7_SH2}.

The partial substituting Tb$^{3+}$ ions with nonmagnetic Y$^{3+}$
ions is naturally expected to have an impact on the
low-temperature magnetism of Tb$_2$Ti$_2$O$_7$. Fig. 7 shows the
temperature dependencies of magnetization with $H \parallel$ [111]
and the zero-field specific heat of Tb$_2$Ti$_2$O$_7$,
Tb$_{1.8}$Y$_{0.2}$Ti$_2$O$_7$, and TbYTi$_2$O$_7$ crystals. It is
found that the magnetic properties of Tb$_{2-x}$Y$_{x}$Ti$_2$O$_7$
is not a simple combination of the spin liquid Tb$_2$Ti$_2$O$_7$
and nonmagnetic material Y$_2$Ti$_2$O$_7$. For the Y-doping effect
on the specific heat, the magnitudes of the two peaks decrease and
the 6 K peak is almost completely suppressed when the Y$^{3+}$
content reaches 50 \%, which strongly indicates that the
nonmagnetic Y$^{3+}$ doped in Tb$_2$Ti$_2$O$_7$ weakens the
magnetic interaction of Tb$^{3+}$. Moreover, the 0.7 K peak moves
to higher temperature with doping, which indicates that the
Schottky anomaly related ground state doublet splitting is
enlarged by the partial Y$^{3+}$ doping. This could be related to
the structural distortions induced by Y doping. In general, the
substitution of Y$^{3+}$ ions for Tb$^{3+}$ ions only results in a
moderate effect on the magnetism of Tb$_2$Ti$_2$O$_7$. These
results are essentially consistent with those from Muon spin
relaxation and neutron spin echo measurements, which have revealed
that Y$^{3+}$ doping slows down the spin fluctuations but the
cooperative paramagnetic behavior still persists down to very low
temperatures \cite{TbYTi2O7}.

\section{CONCLUSIONS}

High quality and large $R_2$Ti$_2$O$_7$ ($R$ = Gd, Tb, Dy, Ho, Y,
Er, Yb and Lu) and Tb$_{2-x}$Y$_x$Ti$_2$O$_7$ ($x$ = 0.2 and 1)
single crystals were grown by the optical floating-zone method.
The growth conditions were optimized and were found to be
dependent on the radius of the $R^{3+}$ ions. The structure and
crystallinity of single crystals were characterized by X-ray
diffraction and Laue photographs. The physical properties were
characterized by low-tempearture magnetization and specific heat
measurements.

\section*{ACKNOWLEDGMENTS}

This work was supported by the National Natural Science Foundation
of China, the National Basic Research Program of China (Grant Nos.
2009CB929502 and 2011CBA00111), and the Fundamental Research Funds
for the Central Universities (Program No. WK2340000035).






\bibliographystyle{elsarticle-num}
\bibliography{<your-bib-database>}





\newpage

\begin{figure}[tp]
\vglue 1.0cm
\newpage


\caption{(Color online) Single crystals of Dy$_2$Ti$_2$O$_7$
grown at 2.5 mm/h under 0.3 MPa O$_2$ pressure (a) and at 4 mm/h
under 0.4 MPa O$_2$ pressure (b), respectively. (c) Single crystal
shown in panel (b) after annealing at 900$^\circ$C in
normal-pressure O$_2$.}


\caption{X-ray diffraction pattern of (110) plane (a) and the
rocking curve of (440) peak (b) for a piece of Dy$_2$Ti$_2$O$_7$
crystal, which was orientated by using the X-ray Laue
photographs.}


\caption{(Color online) Single crystals of Ho$_2$Ti$_2$O$_7$ (a),
Tb$_2$Ti$_2$O$_7$ (b), and Gd$_2$Ti$_2$O$_7$ (c) grown under 0.4
MPa O$_2$ pressure at rates of 4 mm/h, 4 mm/h, and 2.5 mm/h,
respectively. Single crystals of Er$_2$Ti$_2$O$_7$ (d) and
Yb$_2$Ti$_2$O$_7$ (e) grown under 0.1 MPa O$_2$ pressure at rates
of 4 mm/h and 3 mm/h, respectively. (f) Single crystal of
Y$_2$Ti$_2$O$_7$ grown at 4 mm/h in O$_2$ and Ar mixture with the
ratio of 4:1.}


\caption{(Color online) Magnetization curves of Gd$_2$Ti$_2$O$_7$
(a) and Er$_2$Ti$_2$O$_7$ (b) single crystals with magnetic field
along the [100], [111], and [110] axes at 2 K. The demagnetization
field is not corrected.}

\caption{(Color online) Magnetic susceptibilies of one
Tb$_2$Ti$_2$O$_7$ single crystal before and after annealed in 0,1
MPa O$_2$. The magnetic fields are applied along the [110] axe.
Inset: the magntizations curves at 2 K.}

\caption{(Color online) Temperature dependencies of specific heat
of $R_2$Ti$_2$O$_7$ ($R$ = Dy, Tb, Gd, Er, Yb, Y and Lu) single
crystals in zero field.}

\caption{(Color online) Temperature dependencies of magnetization
with magnetic field along the [111] axes (a) and the specific heat
(b) of Tb$_2$Ti$_2$O$_7$, Tb$_{1.8}$Y$_{0.2}$Ti$_2$O$_7$, and
TbYTi$_2$O$_7$ single crystals. The data is calculated with per
mole of formula unit.}

\end{figure}
\clearpage

\begin{figure*}[htbp]
\center {$\Huge\textbf{Fig. 1} $}
\includegraphics[bb = 10 800 700 500, width=1.0\textwidth]{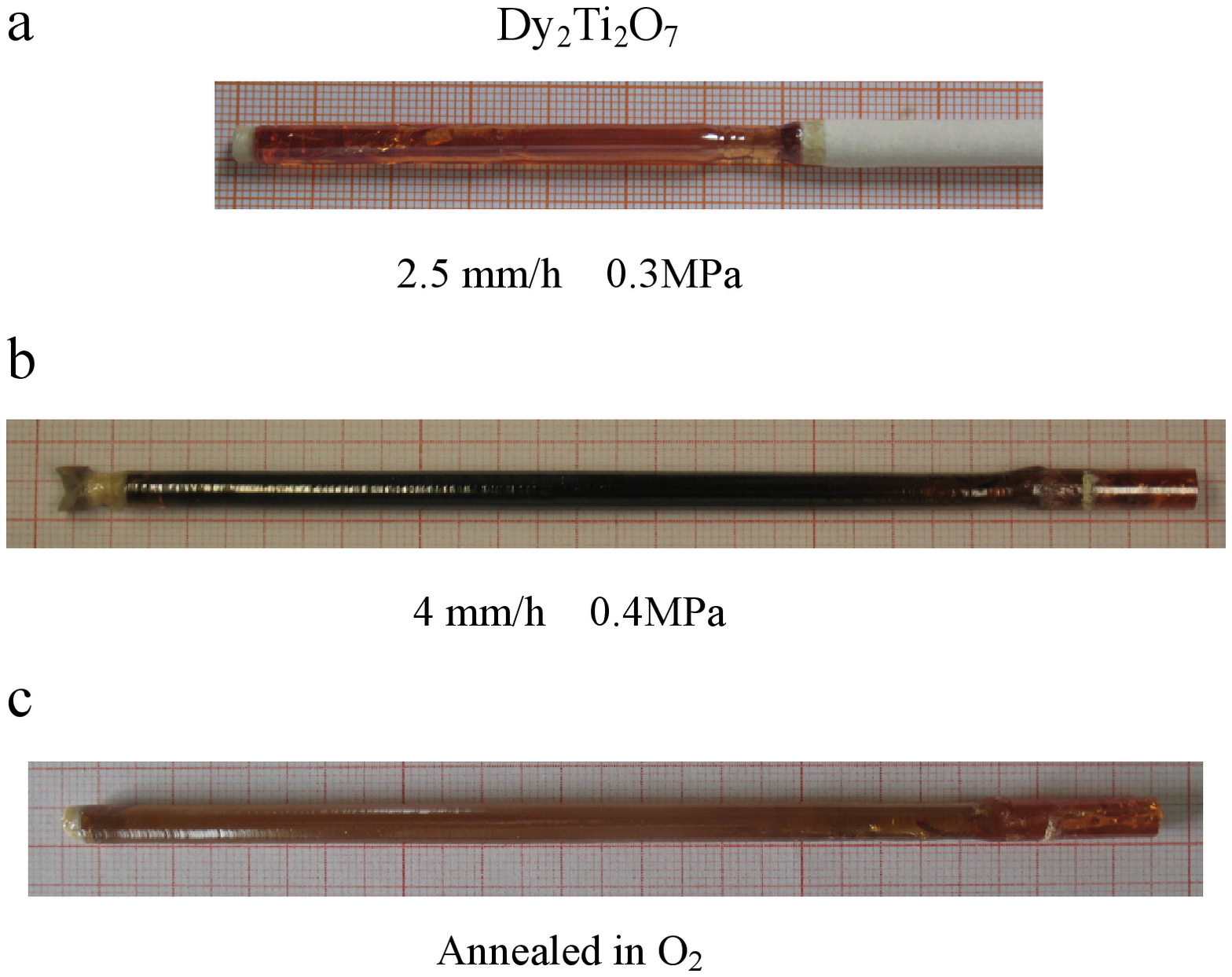}
\end{figure*}
\clearpage
\newpage

\clearpage
\newpage
\begin{figure*}[htbp]
\center {$\Huge\textbf{Fig. 2} $}

\includegraphics[bb = 10 800 700 500, width=1.2\textwidth]{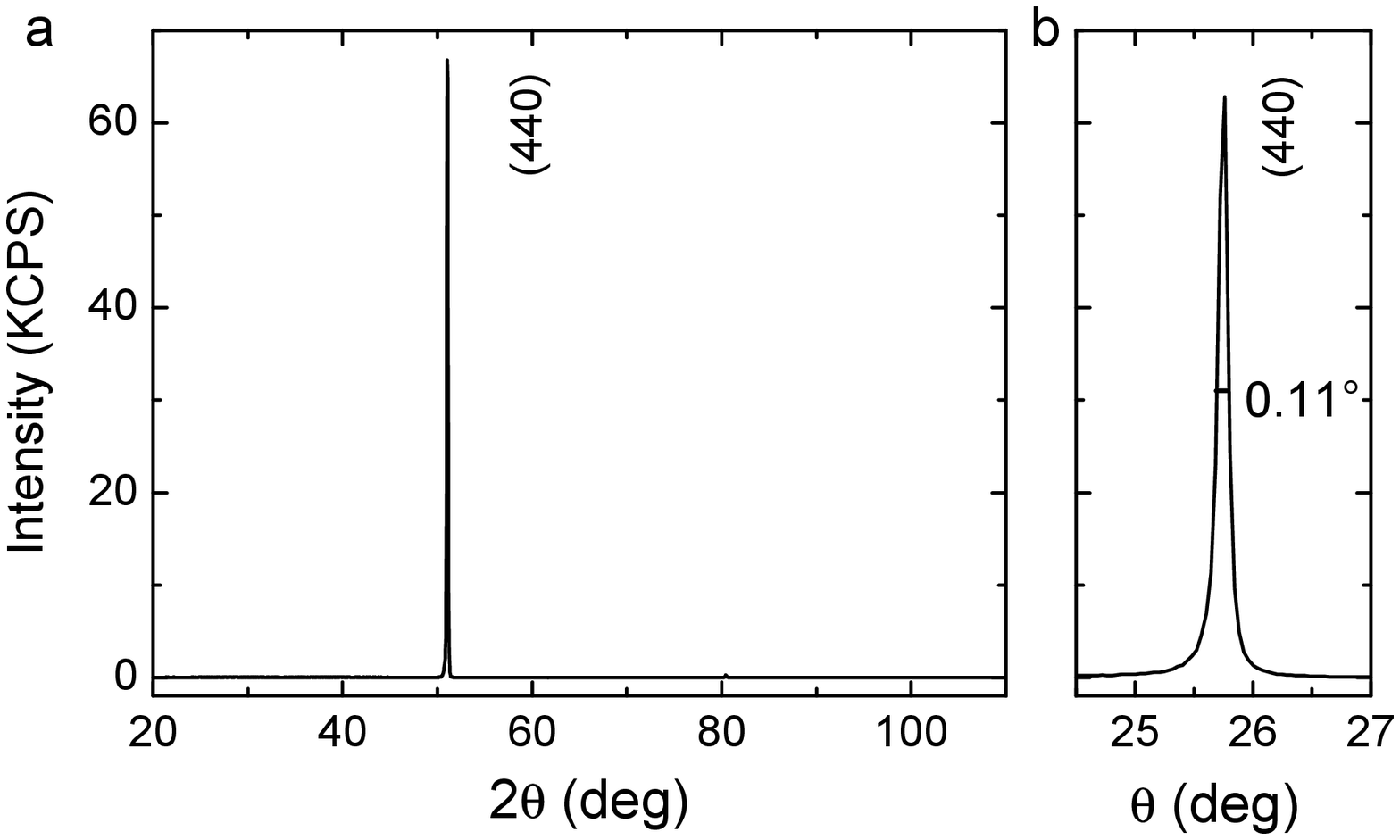}
\end{figure*}
\clearpage
\newpage

\begin{figure*}[htbp]
\center {$\Huge\textbf{Fig. 3} $}

\includegraphics[bb = 10 800 700 500, width=1.2\textwidth]{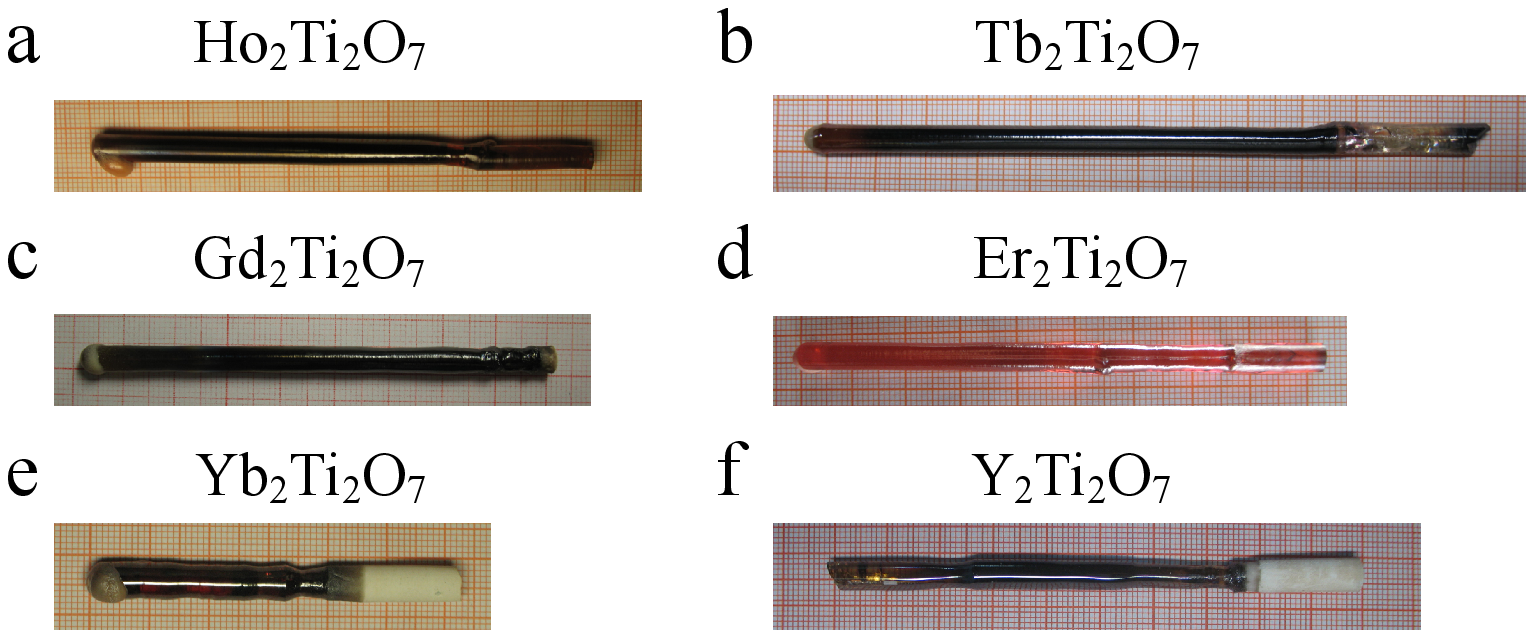}
\end{figure*}
\clearpage
\newpage

\clearpage
\newpage
\begin{table}[htbp]
\caption {Optimized growth conditions for different rare-earth
titanates. The lattice parameters are obtained from the powder
X-ray diffraction of single crystals.} \centering
\begin{tabular}{p{75pt}p{60pt}p{95pt}p{75pt}p{120pt}}
\hline  \hline   & R$^{3+}$ radius\ ({\AA}) \centering & Lattice
parameter ({\AA}) \centering & Growth rate (mm/h) \centering &
Atomosphere\\\hline
  Gd$_2$Ti$_2$O$_7$ & 0.94 & 10.196(9) & 2.5 & 0.4 MPa O$_2$\\
  Tb$_2$Ti$_2$O$_7$ & 0.92 & 10.155(2) & 2.5, 4 & 0.4 MPa O$_2$\\
  Dy$_2$Ti$_2$O$_7$ & 0.91 & 10.127(5) & 4 & 0.4 MPa O$_2$\\
  Ho$_2$Ti$_2$O$_7$ & 0.89 & 10.105(2) & 4 & 0.4 MPa O$_2$\\
  Y$_2$Ti$_2$O$_7$ & 0.89 & 10.089(3) & 4 & 0.1 MPa O$_2$+Ar(4:1)\\
  Er$_2$Ti$_2$O$_7$ & 0.88 & 10.072(3) & 4 & 0.1 MPa O$_2$\\
  Yb$_2$Ti$_2$O$_7$ & 0.88 & 10.033(2) & 3 & 0.1 MPa O$_2$\\
\hline \hline
\end{tabular}
\end{table}
\clearpage
\newpage

\begin{figure*}[htbp]
\center {$\Huge\textbf{Fig. 4} $}

\includegraphics[bb = 10 800 700 700, width=1.0\textwidth]{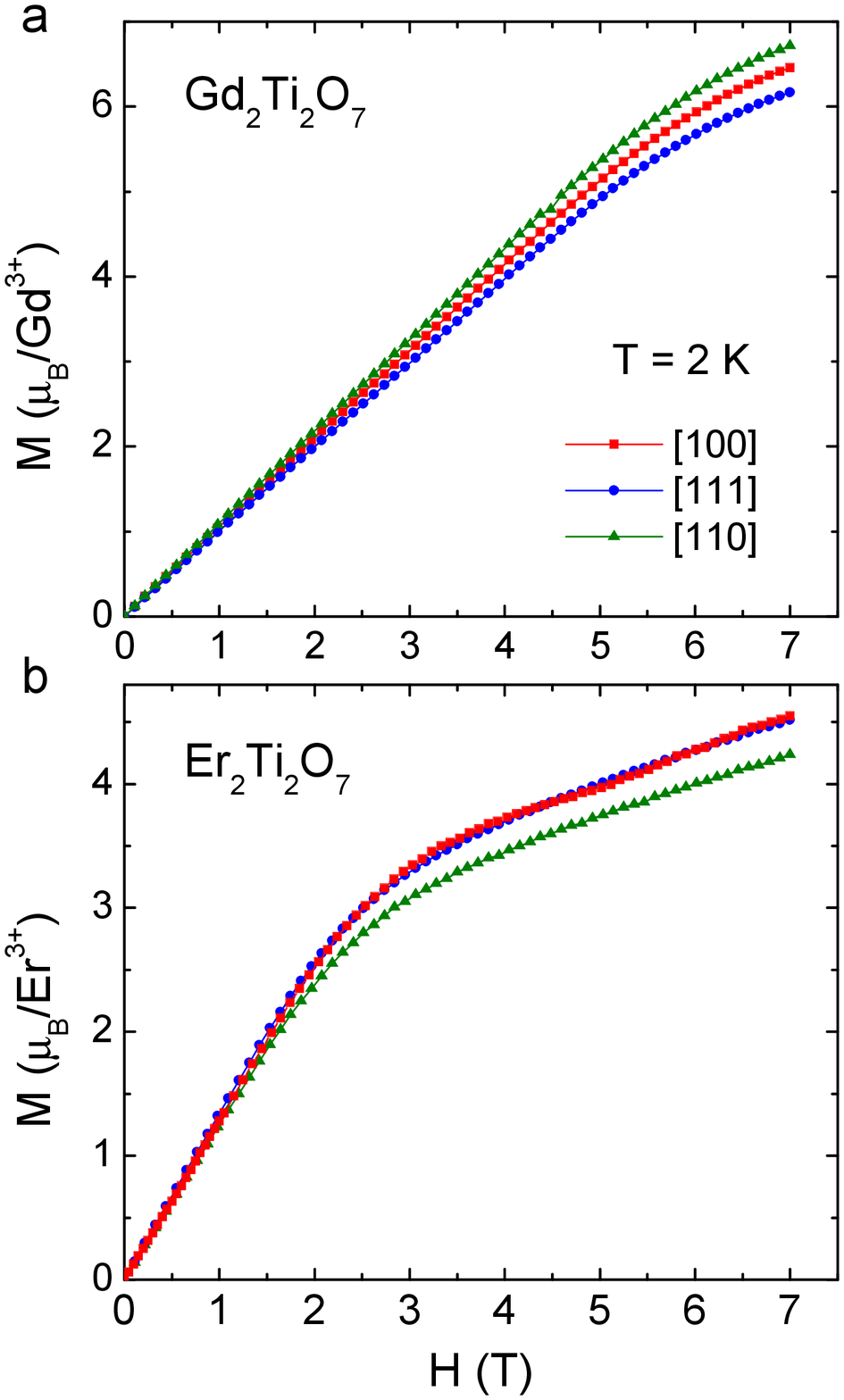}
\end{figure*}
\clearpage
\newpage

\begin{figure*}[htbp]
\center {$\Huge\textbf{Fig. 5} $}

\includegraphics[bb = 10 800 700 700, width=0.8\textwidth]{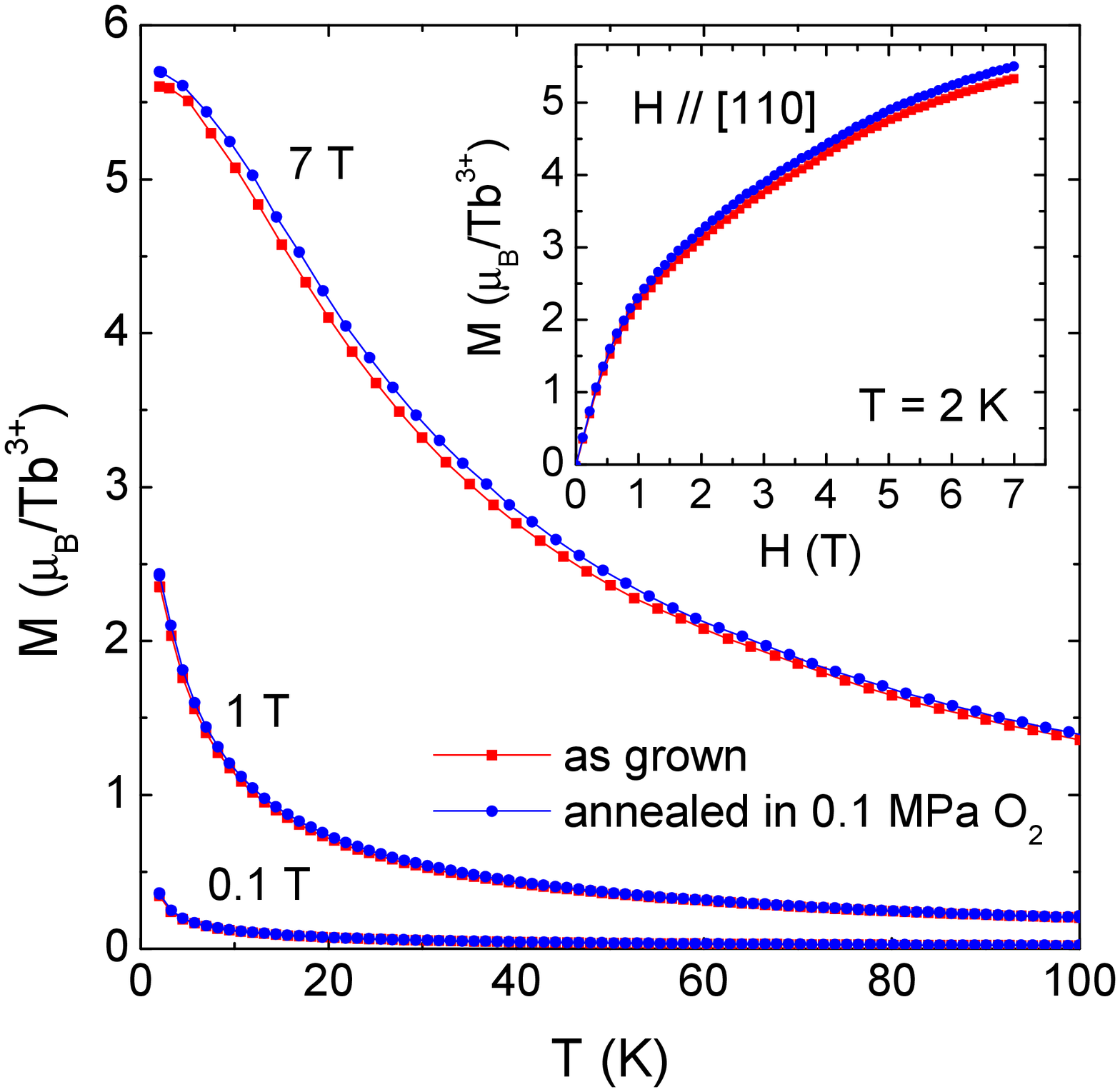}
\end{figure*}
\clearpage
\newpage

\begin{figure*}[htbp]
\center {$\Huge\textbf{Fig. 6} $}

\includegraphics[bb = 10 800 700 700, width=0.8\textwidth]{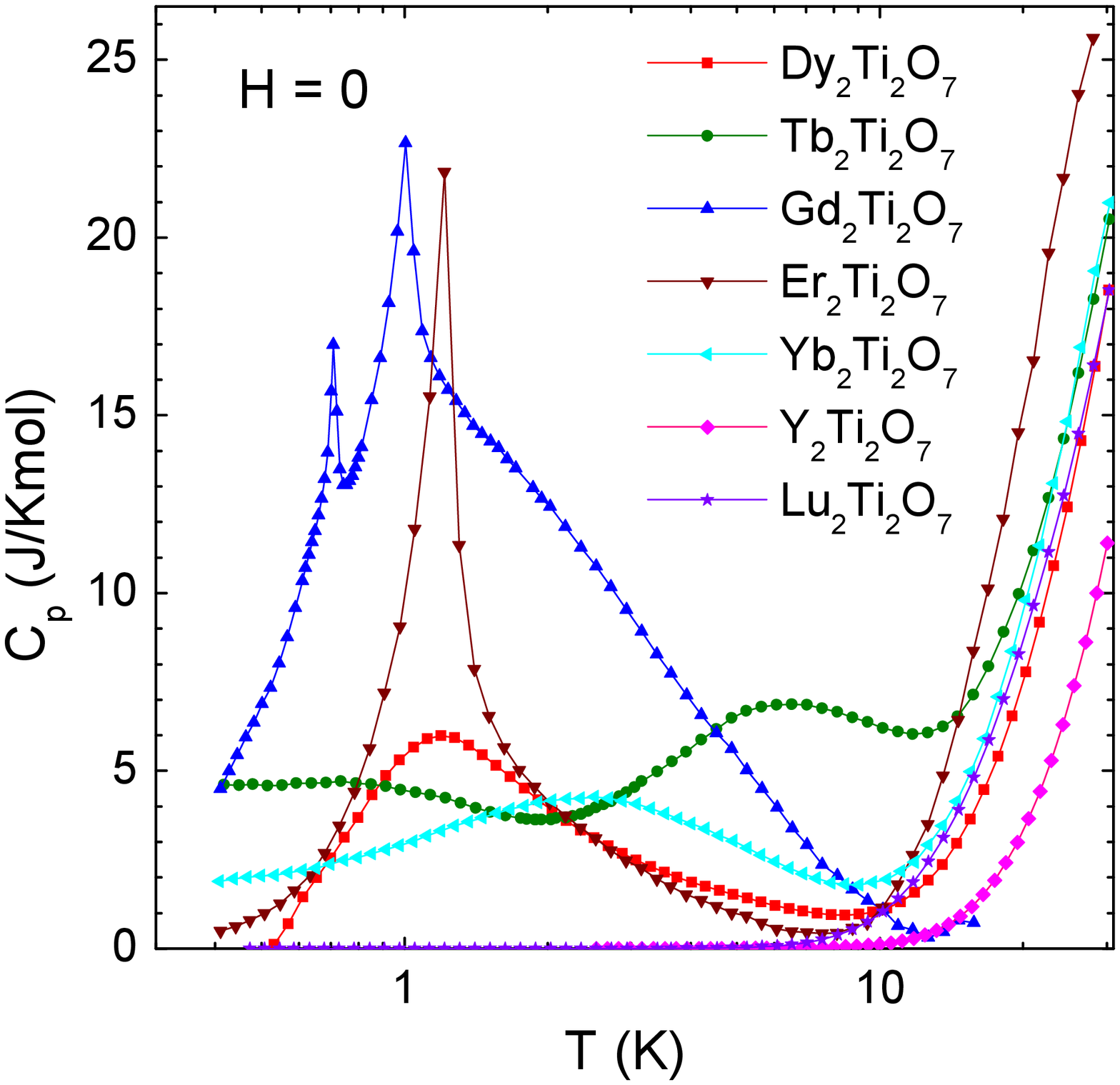}
\end{figure*}
\clearpage
\newpage

\begin{figure*}[htbp]
\center {$\Huge\textbf{Fig. 7} $}

\includegraphics[bb = 10 800 700 700, width=1.0\textwidth]{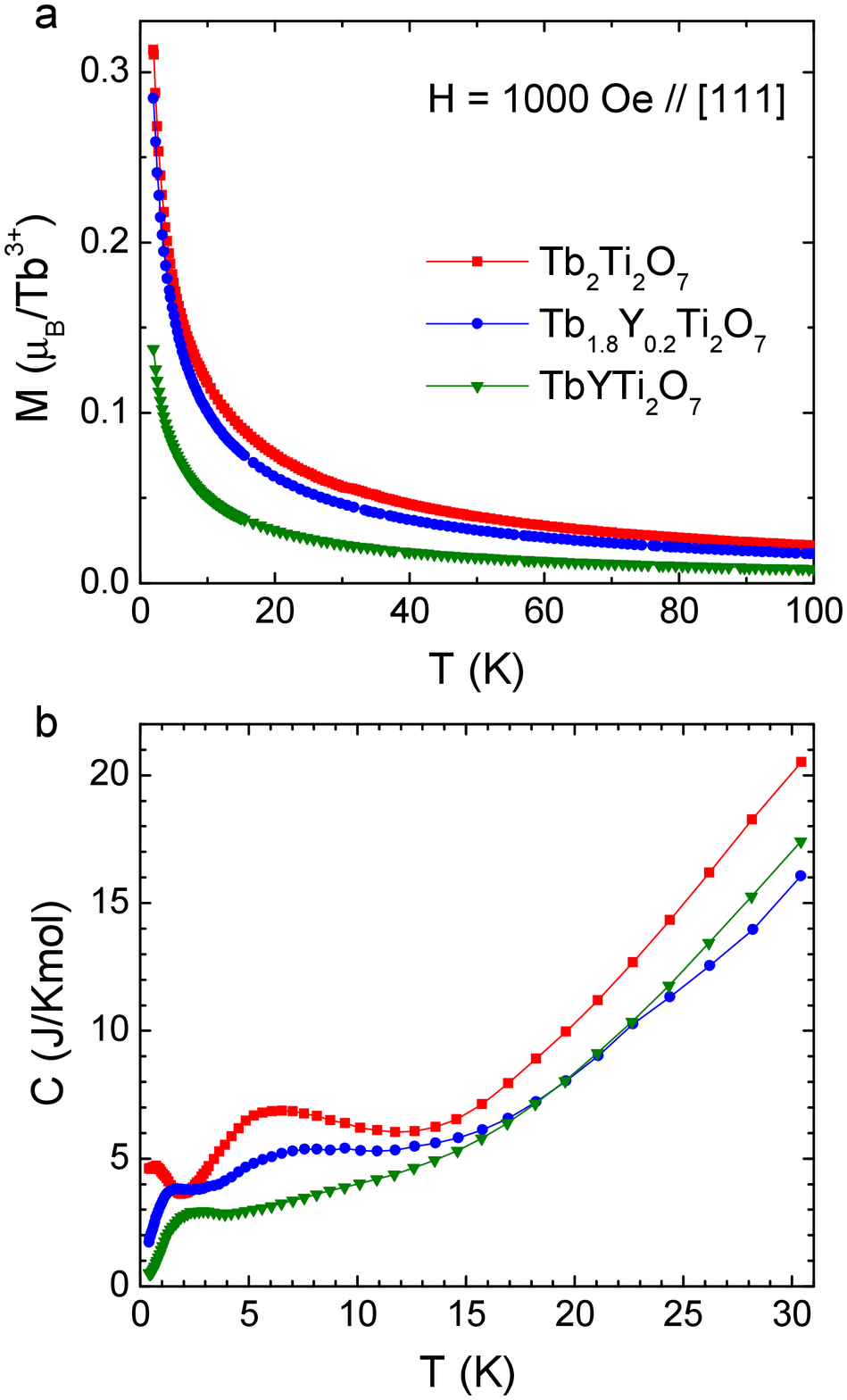}
\end{figure*}
\clearpage
\newpage

\end{document}